%% file: RLS_Precoder_RA_TA.tex
\documentclass[conference]{IEEEtran}

\usepackage{amsmath,dsfont,bbm,epsfig,amssymb,amsfonts,amstext,verbatim,amsopn,cite,subfigure}
\usepackage{multirow,multicol,lipsum,xfrac}
\usepackage{amsthm}
\usepackage{mathtools,amsthm}
\usepackage{url}
\usepackage{amsfonts}
\usepackage{epsfig}
\usepackage[font=small]{caption}
\usepackage{psfrag}	
\usepackage[Algorithm,ruled]{algorithm}
\usepackage{algorithmicx}
\usepackage{algpseudocode}
\usepackage{pifont}
\usepackage[utf8]{inputenc}
\usepackage[T1]{fontenc}  
\usepackage[nolist]{acronym}
\usepackage{paralist}
\usepackage{enumitem}
\usepackage{bbm}
\usepackage[process=auto]{pstool}
\usepackage{tikz,pgfplots}
\usetikzlibrary{shapes.misc,arrows,positioning}

\include{commands}

\begin{document}

\begin{acronym}
\acro{mimo}[MIMO]{multiple-input multiple-output}
\acro{csi}[CSI]{channel state information}
\acro{awgn}[AWGN]{additive white Gaussian noise}
\acro{iid}[i.i.d.]{independent and identically distributed}
\acro{uts}[UTs]{user terminals}
\acro{bs}[BS]{base station}
\acro{tas}[TAS]{transmit antenna selection}
\acro{glse}[GLSE]{generalized least square error}
\acro{rhs}[r.h.s.]{right hand side}
\acro{lhs}[l.h.s.]{left hand side}
\acro{wrt}[w.r.t.]{with respect to}
\acro{mmW}[mmWave]{millimeter-wave}
\acro{np}[NP]{non-deterministic polynomial-time}
\acro{papr}[PAPR]{peak-to-average power ratio}
\acro{rzf}[RZF]{regularized zero forcing}
\acro{snr}[SNR]{signal-to-noise ratio}
\acro{rfc}[RFC]{radio frequency chain}
\acro{mf}[MF]{matched filtering}
\acro{gamp}[GAMP]{generalized AMP}
\acro{amp}[AMP]{approximate message passing}
\acro{vamp}[VAMP]{vector AMP}
\acro{map}[MAP]{maximum-a-posterior}
\acro{mmse}[MMSE]{minimum mean squared error}
\acro{ap}[AP]{average power}
\acro{pa}[PA]{power amplifier}
\acro{tdd}[TDD]{time division duplexing}
\acro{rss}[RSS]{residual sum of squares}
\acro{rls}[RLS]{regularized least-squares}
\acro{dbb}[DBB]{digital base-band}
\acro{ra}[RA]{reflect-array}
\acro{ta}[TA]{transmit-array}
\acro{had}[HAD]{hybrid analog-digital}
\end{acronym}

\title{PAPR-Limited Precoding in Massive MIMO Systems with Reflect- and Transmit-Array Antennas}

\author{
\IEEEauthorblockN{
Ali Bereyhi\IEEEauthorrefmark{1},
Vahid Jamali\IEEEauthorrefmark{1},
Ralf R. M\"uller\IEEEauthorrefmark{1},
Georg Fischer\IEEEauthorrefmark{1},
Robert Schober\IEEEauthorrefmark{1}, and
Antonia M. Tulino\IEEEauthorrefmark{2}
}
\IEEEauthorblockA{
\IEEEauthorrefmark{1}Friedrich-Alexander Universit\"at Erlangen-N\"urnberg,\\
\IEEEauthorrefmark{2}Nokia Bell Labs and University degli Studi di Napoli Federico II\\
\{ali.bereyhi, vahid.jamali, ralf.r.mueller, georg.fischer, robert.schober\}@fau.de, a.tulino@nokia-bell-labs.com
\thanks{This work has been presented in the 2019 Asilomar Conference on Signals, Systems, and Computers. The link to the final version in the proceedings will be available later.}
}
}


\IEEEoverridecommandlockouts

\maketitle

\begin{abstract}
Conventional hybrid analog-digital architectures for millimeter-wave massive multiple-input multiple-output (MIMO) systems suffer from poor scalability and high implementational costs. The former is caused by the high power loss in~the~analog network, and the latter is due to the fact that classic MIMO~transmission techniques require power amplifiers with high back-offs.

This paper proposes a novel hybrid analog-digital architecture which addresses both of these challenges. This architecture~implements the analog front-end via a passive reflect- or transmit-array to resolve the scalability issue. To keep the system cost-efficient, a digital precoder is designed whose peak-to-average power ratio (PAPR) on each active antenna is tunable. Using the approximate message passing algorithm, this precoder is implemented with tractable computational complexity. The proposed architecture allows for the use of power amplifiers with low back-offs which reduces the overall radio frequency cost of the system.~Numerical results demonstrate that for low PAPRs, significant performance enhancements are achieved compared to the state of the~art.\vspace*{2mm}

%Reflect- and transmit-arrays offer promising massive MIMO architectures which are energy-efficient and fully scalable in terms of the number of transmit antennas. This paper proposes a hybrid analog-digital precoding scheme for these architectures in which the peak-to-average power ratio on each active antenna is tunable. Using the approximate message passing algorithm, the precoder is implemented with tractable computational complexity. Numerical results demonstrate that for low peak-to-average power ratios, significant performance enhancements are achieved compared to the state of the~art.
\end{abstract}


\IEEEpeerreviewmaketitle
%\vspace*{-3mm}
\section{Introduction}
\label{sec:intro}
Recent studies have shown the necessity of~employing~massive \ac{mimo} settings and moving towards the \ac{mmW} spectrum to meet the data rate demands in the next generations of~wireless~communication systems \cite{rappaport2013millimeter,delgado2018feasible}. This fact has drawn attention to the concept of \ac{had} precoding~\cite{el2014spatially,molisch2017hybrid}. Conventional \ac{had} transmitters suffer from high power loss in their analog~feed~networks. As a solution to this issue, an energy-efficient \ac{had} architecture based~on~\ac{ra} and \ac{ta} antennas has been recently proposed in \cite{jamali2019scale}. In contrast to conventional designs,~this~architecture~is \textit{fully scalable} with respect to the number of antennas. %in Fig.~\ref{fig:1} and reviewed in~Section~\ref{sec:sys}. 

The initial design in \cite{jamali2019scale} employs linear precoding in~the~digital unit. Despite the advantage of low computational complexity, there is a downside to such precoders: \textit{the \ac{papr} at each \ac{rfc} is not restricted}. Noting that the implementation cost scales with the dynamic range\footnote{By dynamic range, we mean the power range in which the power amplifier of the \ac{rfc} behaves linearly. This range is often quantified by the back-off~of the power amplifier.} of the \ac{rfc}s, high \ac{papr} either increases the cost for a desired performance level or causes performance degradation for a fixed budget.

\subsection{Contributions}
In this study, we address the problem of constraining the \ac{papr} of \ac{had} architectures focusing on the recent~proposal in \cite{jamali2019scale}. Invoking the \ac{glse}~scheme introduced in \cite{bereyhi2017nonlinear,bereyhi2017asymptotics,bereyhi2018glse}, we design a digital precoder whose~output \ac{papr} on each active antenna is tunable. The proposed precoder is implemented via the \ac{amp} algorithm, so that its computational complexity scales linearly with the number of active antennas.~Our~results~de-monstrate that by using the proposed scheme the performance is significantly enhanced for low \ac{papr}s.

\subsection{Notation}
Throughout the paper, scalars, vectors, and matrices~are~represented by non-bold, bold lower case, and bold upper case letters, respectively. $\mI_K$ is the $K \times K$ identity matrix,~and~$\mH^{\her}$~is the conjugate transpose of $\mH$. $\norm{\mH}_F$ denotes the Frobenius norm of $\mH$. $\setR$ and $\setC$ are the real axis and complex plane, respectively. $\Ex{\cdot}{}$ denotes mathematical expectation. For simplicity, $\set{1, \ldots , N}$~is abbreviated by $[N]$. %For any differentiable function $\bff(\bx)=\left[ f_1(\bx), \ldots, f_n(\bx) \right]^\trp$,~the~gradient is defined as $\gred{\bx} \bff(\bx) \coloneqq [ \gred{\bx} f_1(\bx), \ldots,\gred{\bx} f_n(\bx) ]^\trp$. 

\section{Problem Formulation}
\label{sec:sys}
Consider downlink transmission in a multiuser \ac{mimo} system with a \ac{bs} and $K$ single-antenna users. The \ac{bs} employs the \ac{ra}/\ac{ta} \ac{had} architecture proposed in \cite{jamali2019scale}.~The architecture is shown in Fig.~\ref{fig:1} and reviewed below.

\subsection{Transceiver Architecture}
\begin{figure}[t]
\centering
\input{Figs/SysModel.tex}
\caption{Block diagram of the \ac{ra}/\ac{ta} \ac{had} transceiver.\vspace*{-6mm}}
\label{fig:1}
\end{figure}
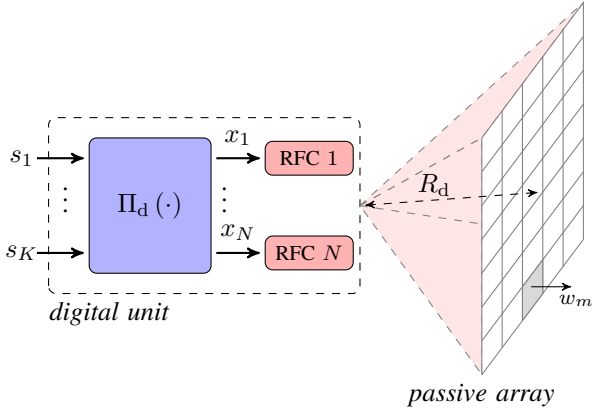

The transmitter consists of a digital signal processing unit with $N$ \ac{rfc}s and a passive array with $M$ antenna elements. This passive array is either an \ac{ra} consisting of $M$~passive~reflectors or a \ac{ta} with $M$ passive re-transmitters. The array is located at distance $R_{\rm d}$ from the \ac{rfc}s; see Fig.~\ref{fig:1}. 

\subsection{HAD Precoding}
Let $s_k$ be the information symbol intended for user $k$. The digital~unit~maps $\bs = \dbc{ s_1,\ldots,s_K }^\trp$ to the transmit signal $\bx\in\setC^{N}$~using~the digital precoder $\Pi_{\rm d} \brc\cdot: \setC^K \mapsto\setC^N$. Hence, the digitally precoded signal is given by
\begin{align}
\bx = \dbb{\bs}.
\end{align}

The transmit signal $\bx$ is then radiated via the \ac{rfc}s towards the passive array. Each element of the array either reflects or re-transmits its receive signal after applying a phase shift.~Following the characterization in \cite{jamali2019scale}, the signal transmitted by the passive array is written as 
\begin{align}
\bw = \mD \mT \bx. \label{eq:xx}
\end{align}
In \eqref{eq:xx}, the matrices $\mT$ and $\mD$ are defined as follows:
\begin{itemize}
\item $\mT\in\setC^{M\times N}$ models the linear channel from the~\ac{rfc}s to~the passive array. In general, $\mT$ depends~on~the~array\\ positioning and characteristics. When all the active antennas radiate with the same pattern and the passive elements are isotropic, the $\brc{m,n}$ entry of $\mT$ is given by~\cite{jamali2019scale}
\begin{align}
\dbc{\mT}_{mn} \hspace*{-.5mm}= \left[ \frac{\lambda \sqrt{ \zeta G\brc{\theta_{mn} , \phi_{mn}}  }}{4 \pi r_{m n } } \exp\set{ - \rmj \frac{2\pi r_{mn} }{\lambda} } \right], \label{eq:T}
\end{align}
for $n\in\dbc{N}$ and $m\in\dbc{M}$. In \eqref{eq:T}, 
\begin{itemize}
\item $\lambda$ is the wavelength.
\item $\zeta$ is the power efficiency of the passive antenna array.
\item $G\brc{\theta , \phi}$ denotes the radiation pattern of the~active~antennas where $\theta$ and $\phi$ are the elevation and~azimuth~angles, respectively.
\item $\brc{r_{mn}, \theta_{mn}, \phi_{mn}}$ is the relative spherical coordinate of the $m$-th passive element when the origin is located at the $n$-th active antenna.
\end{itemize}

\item $\mD$ is an ${M\times M}$ diagonal matrix which~models~the~phase shifts applied by the passive antenna elements. The diagonal entries of $\mD$ are given by
\begin{align}
\dbc{\mD}_{m m} = \exp\set{ \rmj 2\pi \beta_m},
\end{align}
where $0\leq \beta_m \leq 1$. In practice, the~passive~elements~apply quantized phase shifts; for example,~architectures~implemented by transmission lines \cite{chou2018all,abdelrahman2017analysis}. Hence, $\beta_m$ takes values from some set $\setB =\set{B_1,\ldots,B_Q} \subset \dbc{0,1}$, where $Q$ denotes the number of quantization levels.
\end{itemize}

\subsection{Channel Model}
The signal radiated by the passive array is transmitted over a Gaussian broadcast channel which experiences quasi-static fading. The matrix of channel gains is denoted by $\mH\in\setC^{K\times M}$ and is known at the \ac{bs} prior to signal transmission. The vector of receive signals,~i.e., $\by = \dbc{y_1,\ldots,y_K}^\trp$ is hence given by 
\begin{align}
\by = \mH \bw + \bz = \mH  \mD \mT \bx + \bz,
\end{align}
where $\bz$ is additive white Gaussian noise  with zero mean and variance $\sigma^2$, i.e., $\bz\sim\mathcal{CN}\brc{0,\sigma^2 \mI_K}$.

\subsection{Main Objectives}
The \ac{had} transmitter involves $\mT$, $\mD$ and $\Pi_{\rm d}\brc\cdot$, which need to be designed:%has three blocks whose design should be discussed:
\begin{enumerate}
\item[(a)] The design of $\mT$ mainly requires the tuning of the~antenna characteristics and the array position.
\item[(b)] For the design of $\mD$, the optimal phase shift of each~antenna element should be determined.
\item[(c)] The digital precoder $\Pi_{\rm d}\brc{\cdot}$, which is implemented in the baseband domain, needs to be designed~for~a~given~set of signal constraints, e.g. limited transmit~power.
\end{enumerate}

The update rates of these three blocks are not the same.~$\mD$ and $\Pi_{\rm d}\brc\cdot$ are often updated multiple times within a coherence time interval while $\mT$ is designed offline. We hence assume that $\mT$ has been tuned in advance and is kept fixed afterwards.

%\subsection{Main Objective and Performance Measures}
Existing proposals consider linear digital precoders \cite{jamali2019scale},~i.e., $\Pi_{\rm d}\brc\bs = \mA \bs$ for some precoding matrix~$\mA$. Such schemes can result in high \ac{papr} at the \ac{rfc}s. To address this issue, in this paper,~we invoke the \ac{glse} framework, recently introduced in \cite{bereyhi2017nonlinear,bereyhi2017asymptotics,bereyhi2018glse}, and propose a \ac{papr}-limited digital precoder based on the \ac{rls} method. For the sake~of tractability, we further develop an iterative algorithm based~on \ac{amp} to implement the digital precoder. The complexity~of~this algorithm scales linearly with the number of active antenna elements. This makes the proposed scheme practically feasible for massive \ac{mimo} settings. The tuning strategy for~$\mD$~is~further discussed briefly in Section~\ref{sec:analog}.

\subsection{Performance Metric}
To quantify the performance, we consider~the \ac{rss} at the receiver side as the metric. For a given vector of information symbols $\bs$ and its corresponding digitally precoded signal $\bx = \dbb{\bs}$, the \ac{rss} is defined as
\begin{align}
\rss \brc{\bx,\mD \vert \mT} = \norm{ \mH \mD \mT \bx  - \bs  }^2.
\end{align}
The \ac{rss} quantifies the average distortion between~the~noise-free receive signals and the information symbols\footnote{More generally, the \ac{rss} can be defined as the distance between the~noise-free receive signals and a scaled version of the information~symbols,~i.e.,~replacing $\bs$ with $\alpha\bs$ for some scalar $\alpha$. For sake of simplicity, we set $\alpha=1$.}. In~the~ideal case, where the end-to-end channel is inverted,  the \ac{rss} is~zero.%$\rss \brc{\bv,\mD\vert \mT} = 0$.
%which reads% defined as 
%we require a measure of performance with respect to which the precoder is designed. Following \cite{bereyhi2018glse}, we consider
%\begin{definition}[RSS]
%\label{def:RSS}
%Consider the vector of information symbols $\bs$ and let $\bx\in\setC^N$ be its digitally precoded signal. For a given $\mT$, the \ac{rss} at the user terminals is defined as
%\begin{align}
%\end{align}
%\end{definition}

\section{HAD Precoding with Minimum RSS}
For given $\bs$, $\mT$, and signal constraints, the~optimal~choices of $\mD$ and $\Pi_{\rm d} \brc\cdot$ with respect to the \ac{rss} are
\begin{align}
\brc{\Pi_{\rm d} \brc{\bs}, \mD} = 
 \argmin_{\tilde\Pi \in \maP , \tilde\mD \in \maD} \left. \rss \brc{ \tilde{\Pi} \brc{\bs},\tilde\mD\vert \mT} \right. , \label{eq:Optim} 
\end{align}
where $\maP$ is the set of all mappings whose output entries satisfy the given signal constraints. For example, if the transmit signal is restricted to have a limited peak power, $\maP$ contains~all~func-tions $\tilde{\Pi}\brc\cdot:\setC^K \mapsto \setC^N$ for which the entries of the precoded signal $\tilde\bx = \tilde{\Pi}\brc{\bs}$ satisfy $\abs{\tilde{x}_n}^2 \leq P$ for some $P$. Moreover, set $\maD$ contains all possible phase shift matrices. 

In \eqref{eq:Optim}, $\mT$ is specified by the position and characteristics of the passive and active antenna arrays. In this respect, one can design the system such that
\begin{align}
\mT = \argmin_{\tilde\mT\in\maT } \left. \Ex{
  \rss \brc{ {\Pi}_{\rm d} \brc{\bs},\mD\vert\tilde\mT} }{\mH,\bs} \right. \label{eq:T_Design}
\end{align}
where $\maT$ is the set of all possible channel matrices between the active and passive arrays\footnote{Note that in \eqref{eq:T_Design}, the digital precoder and the matrix of phase shifts in the objective function are given by \eqref{eq:Optim} which are also functions of $\tilde{\mT}$.}. The expectation in \eqref{eq:T_Design} averages the \ac{rss} over all realizations of the channel~coefficients, i.e., $\mH$, and the information symbols. %The formulation in \eqref{eq:Optim} describes a regression problem which is effectively addressed via the \ac{rls} method. In the sequel, we use the \ac{rls} method to design the transmitter modules.

In the sequel, we focus on the design of the digital precoder and the passive phase shifters considering a limited \ac{papr} as the design constraint. To this end, we note that
\begin{itemize}
\item The \ac{rss} is in general a mixed function of $\Pi_{\rm d} \brc\cdot$ and~$\mD$ meaning that \eqref{eq:Optim} cannot be decomposed into two decoupled optimization sub-problems in terms of $\Pi_{\rm d} \brc\cdot$ and~$\mD$.
\item The diagonal entries of $\mD$ do not take values~from~a~convex set. Hence, the global minimum of the optimization problem in \eqref{eq:Optim} is not tractable.
\end{itemize}
To address these issues, we decompose the problem in \eqref{eq:Optim} into two mutually coupled sub-problems. In the first sub-problem, we find the optimal precoder as a function~of~$\mD$~via~the~\ac{rls} method. Then, we substitute the solution into \eqref{eq:Optim} and find~an approximation of the optimal choice for matrix $\mD$, noting that this latter task is \ac{np}-hard. We discuss our approach in detail in the following sections.

\subsection{Designing the Digital Unit}
\label{sec:digital}
Assuming a fixed $\mD$, one can interprete the effective end-to-end channel $\mH_{\rm e} = \mH \mD\mT$ as a matrix of $N$ \textit{regressors}, and $\bs$ as a vector of $K$ \textit{regressands} assumed to be linearly related to the regressors via $N$ \textit{regression coefficients}. These coefficients are the entries of the digitally precoded signal $\bx$. By~this~interpretation, \ac{rss} minimization is mathematically equivalent to the least-squares formulation of this linear regression problem\footnote{See \cite{bereyhi2018glse} for more details on this representation.}.

%The joint optimization problem in \eqref{eq:Optim} is not straightforward. To tackle this problem tractably, 
The standard approach to solving this equivalent regression problem is to utilize the \ac{rls} method in which we minimize a penalized version of the \ac{rss}. The penalty term, often referred to as the \textit{regularization term}, is proportional to the constraints required to be satisfied by the coefficients. Considering the particular constraint of \ac{papr} limitation, we note the following items:
\begin{enumerate}
\item The average power of the transmit signal is required~to~be restricted. Such a constraint can be imposed by penalizing the \ac{rss} with a term proportional to $\norm{\bx}^2$.
\item The power of the entries of $\bx$, i.e., $x_n$ for $n\in\dbc{N}$, should be bounded from above. Such a constraint is enforced by adding a barrier function to the \ac{rss} which tends to infinity as $\abs{x_n}^2 > P$ for some peak~power~$P$.~Alternatively, we can restrict the support of the entries of $\bx$, over which we minimize the \ac{rss}, to 
\begin{align}
\setX = \set{ x \in \setC : \abs{x}^2 \leq P }. \label{eq:X}
\end{align}
\end{enumerate}

\hspace{-.5mm}Considering the above discussion, the \ac{rls}-based digital~pre-coder with limited \ac{papr} is given by
\begin{align}
\Pi_{\rm d} \brc{\bs \vert \mD, \mT, \mH} = \argmin_{\bv \in \setX^N} \ \norm{ \mH \mD \mT \bv  -  \bs  }^2 + \mu_{\rls} \norm{\bv}^2  \label{eq:GLSE} 
\end{align}
with  $\setX$ given in \eqref{eq:X} and some scalar $\mu_{\rls}$. We refer to $\mu_{\rls}$ as the \textit{regularizer}. The arguments  $\mD$, $\mT$, and $\mH$ further indicate the dependency of $\Pi_{\rm d} \brc\cdot$ on the realizations of the~channel~coefficients and analog modules. 

The precoder in \eqref{eq:GLSE} describes a \textit{state-dependent} \ac{glse}~precoder \cite{bereyhi2018glse} whose output is calculated in polynomial time via convex optimization techniques. However, one can~further~reduce the computational complexity via \ac{amp}. In \cite{bereyhi2018precoding}, a class of iterative algorithms based on \ac{amp} has been developed for generic \ac{glse} precoding schemes. Invoking the formulation in \cite{bereyhi2018precoding}, the precoding scheme in \eqref{eq:GLSE} is implemented iteratively, such that its complexity grows linearly with $N$. For the sake of brevity, we skip the derivations for the \ac{amp}-based algorithm and refer interested readers to \cite{bereyhi2018precoding} for more details.
%For detailed discussions, see \cite{bereyhi2018glse}.

\subsection{Designing the Phase Shifters}
\label{sec:analog}
%$\mT$ depends on the array positioning and antenna characteristics. Hence, it is designed offline; see discussions in \cite{jamali2019scale}. 
The digital precoder in \eqref{eq:GLSE} is designed as a function of $\mD$. The phase shifts for the passive antenna elements are hence optimally set by minimizing the \ac{rss} with respect~to~$\mD$. In other words, $\mD$ is optimally tuned by solving %$\mD$ is efficiently tuned by
\begin{align}
\mD = \argmin_{\tilde\mD \in \maD} \left. \rss \brc{ \Pi_{\rm d} \brc{\bs \vert \tilde\mD, \mT, \mH },\tilde\mD \vert \mT} \right. . \label{eq:Dopt}
\end{align}
$\mD$ can be written as $\mD = \mathrm{diag}\set{\bd}$ for some vector $\bd\in\setD^{M}$, where\footnote{In the limiting case of $Q\to \infty$, $\setD$~converges to the unit circle.}
\begin{align}
\setD = \set{ \exp\set{\rmj 2\pi B_1 }, \ldots, \exp\set{\rmj 2\pi B_Q }}
\end{align}
with $0\leq B_q \leq 1$ for $q\in\dbc{Q}$. As a result, the optimal choice for $\mD$ is written as $\mD = \mathrm{diag}\set{\bd^\star}$ where
\begin{align}
\bd^\star = \argmin_{{\bd} \in \setD^M} \left. f\brc{\bd} \right.   \label{eq:Dopt2}
\end{align}
with
\begin{align}
f\brc{\bd} =  \left. \rss \brc{ \Pi_{\rm d} \brc{\bs \vert \mathrm{diag}\set{\bd}, \mT, \mH },\mathrm{diag}\set{\bd} \vert \mT} \right. .  \label{eq:Dopt3}
\end{align}

The direct approach to solving \eqref{eq:Dopt2} is not tractable~for~the following reasons:
\begin{enumerate}
\item Since the precoded signal is a function of $\mD$, $f\brc{\bd}$ does not have a simple analytical form.
\item Even for convex forms of $f\brc{\bd}$, the optimization problem in \eqref{eq:Dopt2} reduces to the problem of integer programming, and hence is an \ac{np}-hard problem\footnote{Note that the problem remains \ac{np}-hard, even when $Q\to \infty$.}.
\end{enumerate}
We hence develop a suboptimal approach which approximates the solution of \eqref{eq:Dopt2}. This approach finds $\mD$ and its corresponding precoded signal iteratively. In each iteration, 
\begin{itemize}
\item the precoded signal~$\bx$~is first determined for a fixed $\mD$ from \eqref{eq:GLSE};
\item $\mD$ is then updated~by \eqref{eq:Dopt2} treating $\bx$ as a constant vector. Note that for a constant $\bx$, $f\brc{\bd}$ has a quadratic~form.
\end{itemize}
These steps are then repeated with the updated version of $\mD$. The algorithm continues iterating until it  either fulfills~a~stop criterion or exceeds the maximum number of iterations. Optimization problem \eqref{eq:Dopt2} is solved in each iteration via~a~suboptimal computationally tractable algorithm. We refer~to~this algorithm as $\mathrm{Alg} \brc\cdot$; an example of $\mathrm{Alg} \brc\cdot$ can be derived via the \textit{iterative gradient projection} technique \cite{tranter2017unitMLS} or \textit{manifold optimization} \cite{absil2009optimization}. For more examples~of such algorithms; see \cite{jamali2019scale,yu2016alternating,ioushua2019family}, and the references therein.

The proposed approach~with~an~exemplary stop criterion is summarized in Algorithm~\ref{A-GAMP}.

%In this algorithm, \texttt{A} is calculated via \ac{amp}, and \texttt{B} is determined via a conventional method available in the literature.

%Noting that for a fixed $\bv$ and $\mT$, $\rss\brc{\bv,\mD,\mT}$ is a quadratic function of $\mD$, it is straightforward to show that the complexity of this algorithm linearly scales with $M$.

\begin{algorithm}[t]
\caption{Iterative HAD Precoding}
\label{A-GAMP}
\begin{algorithmic}[0]
\Initiate Set $\bx_0$ and $\mD_0 $ to some feasible initial values. Choose  threshold  $D_{\rm Th}$ and maximum number of iteration~$T_{\max}$.\vspace*{.5mm}
\While\NoDo $\norm{\mD_{t+1}-\mD_t}_{ F} \geq D_{\rm Th}$ and $t \leq T_{\max}$\vspace*{.5mm}
\begin{itemize}
\item[{$\blacktriangleright$}] Update the precoded signal as
\begin{align*}
\bx_{t+1} = \Pi_{\rm d} \brc{\bs\vert \mD_t, \mT, \mH}
\end{align*}
with $\Pi_{\rm d}\brc\cdot$ given in \eqref{eq:GLSE}.
\item[{$\blacktriangleright$}] Update the phase shifts as
\begin{align*}
\mD_{t+1} = \mathrm{Alg} \brc{ \bx_{t+1},\mH,\mT}
\end{align*}
where $\mathrm{Alg} \brc\cdot$ is an algorithm approximating the solution of \eqref{eq:Dopt2} for
\begin{align*}
f\brc{\bd} =  \left. \rss \brc{ \bx_{t+1} ,\mathrm{diag}\set{\bd} \vert \mT} \right. . 
\end{align*}
\item[{$\blacktriangleright$}] Update $t \leftarrow t+1$.
\end{itemize}
\EndWhile \vspace*{1mm}
\Out $\bx_{T}$ and $\mD_{T}$, where $T$ is index of the final iteration.
\end{algorithmic}
\end{algorithm}

\section{Numerical Results}
To investigate the performance of the proposed scheme, we study a scenario with $K=8$ users, $N=4$~\ac{rfc}s, and $M=64$ passive elements. Throughout the simulations, the wavelength is set to $\lambda = 5$ mm.

\subsection{Settings for the Passive and Active Arrays}
The passive array contains $64$ antenna elements installed on an aperture of size $4\lambda \times 4\lambda$. The active antennas are further arranged on a ring with radius $R_{\rm r} = \lambda$ parallel to the passive array. The center of the ring and the passive array coincide on the horizontal plane and are separated by $R_{\rm d} = 4\lambda / \sqrt{\pi}$.

The active antennas on the \ac{rfc}s radiate with the~same~pattern which is horizontally omnidirectional and vertically uniform over $\dbc{ \pi/6 , 5\pi/6}$. This means that\footnote{This radiation pattern is assumed for sake of simplicity. In practice,~patterns with high directivity towards the passive array can be used in order to avoid unwanted power loss in the system.} %The antenna gain is set to one in the main lobe,~and~zero outside. %
\begin{align}
G\brc{\theta,\phi} =
\begin{cases}
1 &\theta \in \dbc{ \pi/6 , 5\pi/6} \ \text{and} \ \phi \in \dbc{0 , 2\pi}\\
0 & \text{otherwise}
\end{cases}.
\end{align}
%where $\theta$ and $\phi$ denote the elevation and azimuth, respectively. 

\subsection{Channel, Information Symbols, and Transform $\mT$}
The entries of $\mH$, as well as the information symbols, are generated independently, where $\dbc{\mH}_{km}\sim\mathcal{CN}\brc{0, 1/M}$ and $s_k \sim\mathcal{CN}\brc{0, 1}$ for $k\in\dbc{K}$ and $m\in\dbc{M}$. $\mT$ is first calculated from \eqref{eq:T}, and then normalized, such that all the entries lie on the unit circle. 

\subsection{Numerical Simulations}
Since our focus is on \ac{papr} restriction at the digital unit, we set $\mD$ to a fixed matrix and do not iterate further to optimize $\mD$. The phase shifts at the passive array can be further optimized by setting $\mathrm{Alg}\brc\cdot$ in Algorithm~\ref{A-GAMP} to~the~gradient~projection scheme given in \cite[Algorithm~1]{tranter2017unitMLS}. %and iterating for multiple steps.

The simulations are given for $J=10^4$ independent realizations. The precoder in \eqref{eq:GLSE} is further implemented via \ac{amp} following the derivations in \cite{bereyhi2018precoding}. To quantify the performance of the proposed scheme, we define two parameters:
\begin{enumerate}
\item The~\textit{per-antenna \ac{papr}} which for active antenna $n$ reads 
\begin{align}
\mathrm{PAPR}_n = \brc{ { \frac{1}{J} \sum_{j=1}^J \abs{x_n\brc{j}}^2 } }^{-1} \max_{j\in\dbc{J} }\left. \abs{x_n\brc{j}}^2 \right. ,
\end{align}
where $\bx\brc{j}$ is the $j$-th realization of the transmit vector. The \ac{papr} is determined by averaging $\mathrm{PAPR}_n$ over $n$. 
\item The \textit{average \ac{rss}} is defined as
\begin{align}
\overline{ \mathrm{RSS} } = \frac{1}{K} \sum_{j=1}^J \rss\brc{\bx\brc{j},\mD,\mT} .
\end{align}
%, and $x_n\brc{j}$ is its $n$-th entry.
%This parameter determines the average multiuser interference per unit of average transmit power at each user terminal.
\end{enumerate}

\begin{figure}
\centering
\input{Figs/Fig1.tex}
\caption{Average RSS versus average PAPR.\vspace*{-5mm}}
\label{fig:2}
\end{figure}
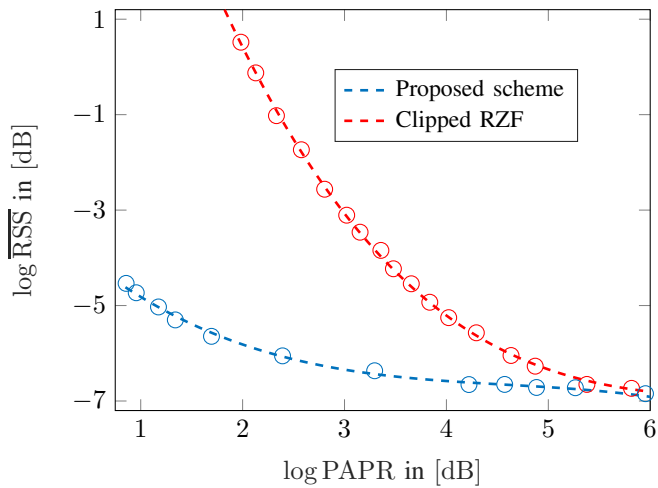

Fig.~\ref{fig:2} shows $\overline{ \mathrm{RSS} }$ versus ${\rm PAPR}$ for the proposed scheme when the peak power is set to $P=1$. The simulation points are found by sweeping the regularizer $\mu_{\rls}$, and the dashed~line is obtained by interpolating the simulation points. From Fig.~\ref{fig:2}, it is observed that as \ac{papr} grows, $\overline{ \mathrm{RSS} }$ decreases. %This follows the fact that for a fixed peak power, the incr both the average transmit power and per-user \ac{rss} reduce, as the \ac{papr} increases. Hence, $\overline{ \mathrm{RSS} }$ can decrease or increase with $\rm PAPR$ depending on the relative change.

For sake of comparison, we further plot the curve for the \ac{rzf} scheme. With \ac{rzf}, the digital precoder reads  %where 
\begin{align}
\tilde\bx = \mH_{\rm e}^\her \brc{ \mH_{\rm e} \mH_{\rm e}^\her + \mu_{\rzf} \mI_K  }^{-1} \bs
%\mA\brc{\mu_\rzf} = \mH_{\rm e}^\her \brc{ \mH_{\rm e} \mH_{\rm e}^\her + \mu_{\rzf} \mI_K  }^{-1}
\end{align}
with $\mH_{\rm e} = \mH \mD \mT$ for some scalar $\mu_{\rm rzf}$. To restrict the peak power, the \ac{rzf} signal is clipped prior to transmission, i.e., 
\begin{align}
x_n = \begin{cases}
\tilde{x}_n &\abs{ \tilde{x}_n}^2 < P \\
\dfrac{ \sqrt{P} \tilde{x}_n }{ \abs{ \tilde{x}_n } } &\abs{ \tilde{x}_n  }^2 > P 
\end{cases},
\end{align}
where $x_n$ denotes the entry transmitted by the $n$-th \ac{rfc}.

As the figure reveals, at low \ac{papr}s, the proposed scheme significantly outperforms \ac{rzf} precoding. This is due to the fact that in this regime, the \ac{rzf} signal is significantly distorted by clipping. For high \ac{papr}, both precoders perform similarly. This observation agrees with the fact that \ac{papr}-limited \ac{glse} precoding reduces to \ac{rzf} as $\mathrm{PAPR}\to \infty$; see \cite{bereyhi2018glse}.

\section{Conclusions and Final Remarks}
The \ac{had} structure proposed in \cite{jamali2019scale} combined with \ac{glse} precoding \cite{bereyhi2018glse} results in a cost-efficient transmitter for massive \ac{mimo} systems. Utilizing \ac{amp}, the preprocessing complexity grows linearly with the number active transmit antennas which is computationally tractable. %More discussions~and further~numerical simulation will be given in the final manuscript.

The framework presented in this paper is also applicable~for other constraints on the transmit signal. Moreover, investigations under realistic millimeter wave channel models further enlighten~the~efficiency of the proposed scheme. The work in these directions is currently ongoing.

\bibliography{ref}
\bibliographystyle{IEEEtran}
\end{document}

%% file: commands.tex
\newcommand{\setX}{\mathbbmss{X}}

\newcommand{\setR}{\mathbbmss{R}}

\newcommand{\setB}{\mathbbmss{B}}

\newcommand{\setC}{\mathbbmss{C}}

\newcommand{\setD}{\mathbbmss{D}}

\newcommand{\Ex}[2]{ \mathbbm{E}_{#2} \left\lbrace #1 \right\rbrace }

\newcommand{\rmj}{\mathrm{j}}

\newcommand{\her}{\mathsf{H}}

\newcommand{\argmin}{\mathop{\mathrm{argmin}}}

\newcommand{\maT}{\mathcal{T}}
\newcommand{\maP}{\mathcal{P}}

\newcommand{\maD}{\mathcal{D}}

\newcommand{\bd}{\mathbf{d}}

\newcommand{\bx}{{\boldsymbol{x}}}

\newcommand{\set}[1]{\left\lbrace#1\right\rbrace}

\newcommand{\brc}[1]{\left( #1 \right) }
\newcommand{\dbb}[1]{\Pi_{\rm d}\left( #1 \right) }
\newcommand{\dbc}[1]{\left[ #1 \right] }

\newcommand{\bz}{{\boldsymbol{z}}}

\newcommand{\bs}{{\boldsymbol{s}}}

\newcommand{\bv}{{\boldsymbol{v}}}

\newcommand{\bw}{{\boldsymbol{w}}}

\newcommand{\by}{{\boldsymbol{y}}}

\newcommand{\trp}{\mathsf{T}}

\newcommand{\mA}{\mathbf{A}}

\newcommand{\mI}{\mathbf{I}}

\newcommand{\mD}{\mathbf{D}}

\newcommand{\mT}{\mathbf{T}}
\newcommand{\mH}{\mathbf{H}}

\newcommand{\rss}{{\mathrm{RSS}}}

\newcommand{\rls}{{\mathrm{rls}}}
\newcommand{\rzf}{{\mathrm{rzf}}}

\newcommand{\norm}[1]{\lVert #1 \rVert}

\newcommand{\abs}[1]{\lvert #1 \rvert}

\newtheoremstyle{mystyle}%                % Name
  {}%                                     % Space above
  {}%                                     % Space below
  {}%                                     % Body font
  {}%                                     % Indent amount
  {\bfseries}%                            % Theorem head font
  {:}%                                     % Punctuation after theorem head
  { }%                                    % Space after theorem head, ' ', or \newline
  {}%                                     % Theorem head spec (can be left empty, meaning `normal')

\theoremstyle{mystyle}

\algnewcommand\algorithmicLet{\textbf{Let}}
\algnewcommand\Let{\item[\algorithmicLet]}
\algnewcommand\algorithmicSet{\textbf{Set}}
\algnewcommand\Set{\item[\algorithmicSet]}

\algnewcommand\algorithmicInitiate{\textbf{Initiate}}
\algnewcommand\Initiate{\item[\algorithmicInitiate]}
\algnewcommand\algorithmicStart{\textbf{Begin}}
\algnewcommand\Begin{\item[\algorithmicStart]}
\algnewcommand\algorithmicEnd{\textbf{End}}
\algnewcommand\End{\item[\algorithmicEnd]}

\algnewcommand\algorithmicOutP{\textbf{Output:}}
\algnewcommand\Out{\item[\algorithmicOutP]}
\newcommand\NoDo{\renewcommand\algorithmicdo{}}

\newcounter{bar}

%% file: Figs/SysModel.tex
\begin{tikzpicture}[scale=.9]

\tikzset{
    %Define standard arrow tip
    >=stealth',
    % Define arrow style
    pil/.style={
           ->,
           thick,
           shorten <=2pt,
           shorten >=2pt,},
    pilak/.style={
           <->,
           thin,
           shorten <=2pt,
           shorten >=2pt,}
}

\draw (2.5,-2.8) node {\textit{passive array}};

\draw [rounded corners=1mm,fill=none,dashed] (-3.9,-1.3) rectangle (.7,1.3) node [midway] { };
\draw (-3,-1.6) node {\textit{digital unit}};

\draw [rounded corners=1mm,fill=blue!30] (-3.3,-1) rectangle (-1.5,1) node [midway] {$\Pi_{\rm d} \brc\cdot$ };

\draw (-4.3,.7) node {$s_1$};
\draw[pil,->] (-4.15,.7) -- (-3.3,0.7);
\draw (-3.65,0.2) node {$\vdots$};

\draw (-4.3,-.7) node {$s_K$};
\draw[pil,->] (-4.15,-.7) -- (-3.3,-0.7);

\draw (-1.1,1) node {$x_1$};
\draw[pil,->] (-1.5,.7) -- (-.7,0.7);
\draw [rounded corners=1mm,fill=red!30] (-.7,.45) rectangle (.6,.95) node[font=\footnotesize,midway] {RFC $1$};

\draw (-1.3,0.2) node {$\vdots$};

\draw (-1.1,-.4) node {$x_N$};
\draw[pil,->] (-1.5,-.7) -- (-.7,-0.7);
\draw[rounded corners=1mm, fill=red!30] (-.7,-.45) rectangle (.6,-.95) node[font=\footnotesize,midway] {RFC $N$};

\filldraw[gray, dashed,fill=red!10] (.7,0) -- (2.5,1) -- (4,3) -- (.7,0);
\filldraw[gray, dashed,fill=red!10] (.7,0) -- (2.5,1) -- (2.5,-2.5) -- (.7,0);
%\filldraw[gray, dashed,fill=red!10] (.7,0) -- (2.5,-2.5) -- (4,-.5) -- (.7,0);
\draw[gray,dashed] (.7,0) -- (4,-0.5);

\draw[gray,fill=white] (2.5,-2.5) -- (2.5,1) --  (4,3) -- (4,-0.5) -- (2.5,-2.5);

\draw[gray] (2.5,-2) -- (4,0);
\draw[gray] (2.5,-1.5) -- (4,0.5);
\draw[gray] (2.5,-1) -- (4,1);
\draw[gray] (2.5,-.5) -- (4,1.5);
\draw[gray] (2.5,0) -- (4,2);
\draw[gray] (2.5,.5) -- (4,2.5);

\draw[gray] (2.8,1.4) -- (2.8,-2.1);
\draw[gray] (3.1,1.8) -- (3.1,-1.7);
\draw[gray] (3.4,2.2) -- (3.4,-1.3);
\draw[gray] (3.7,2.6) -- (3.7,-0.9);
\filldraw[gray,fill=gray!30] (3.1,-1.7) -- (3.1,-1.2) --  (3.4,-.8) -- (3.4,-1.3) -- (3.1,-1.7);

\draw[pilak,dashed,<->] (.7,0) -- (3.4,0.2);
\draw (1.8,.3) node {$R_{\rm d}$};

\draw[pilak,->] (3.15,-1.2) -- (3.9,-1.2);
\draw (3.9,-1.4) node[font=\footnotesize] {$w_m$};

\end{tikzpicture}

%% file: Figs/Fig1.tex
\definecolor{mycolor1}{rgb}{0.00000,0.44700,0.74100}%
\definecolor{mycolor2}{rgb}{0.00000,0.44706,0.74118}%
\begin{tikzpicture}

\begin{axis}[%
width=2.8in,
height=2.1in,
at={(1.976in,1.234in)},
scale only axis,
xmin=0.75,
xmax=6,
xtick={1,2,3,4,5,6},
xticklabels={{$1$},{$2$},{$3$},{$4$},{$5$},{$6$}},
xlabel style={font=\color{white!15!black}},
xlabel={$\log \rm PAPR$ in $ \rm [dB]$},
ymin=-7.2,
ymax=1.2,
ytick={-7,-5,-3,-1,1},
yticklabels={{$-7$},{$-5$},{$-3$},{$-1$},{$1$}},
ylabel style={font=\color{white!15!black}},
ylabel={$\log \overline{ \mathrm{RSS} }$ in $ \rm [dB]$},
axis background/.style={fill=white},
legend style={at={(0.41,.67)}, anchor=south west, legend cell align=left, align=left, draw=white!15!black}
]

\addplot [color=mycolor1, dashed, line width=1.0pt]
  table[row sep=crcr]{%
0.85	-4.61662524244643\\
0.856686385900919	-4.62592119352267\\
0.86763698630137	-4.6410802735534\\
0.88527397260274	-4.66532541681164\\
0.90291095890411	-4.68936166367799\\
0.920547945205479	-4.71319000560929\\
0.938184931506849	-4.73681143406241\\
0.955821917808219	-4.76022694049418\\
0.955945551252391	-4.76039035630068\\
0.973458904109589	-4.78343751636146\\
0.991095890410959	-4.80644415312111\\
1.00137724453597	-4.81976200620532\\
1.00873287671233	-4.82924784222996\\
1.0263698630137	-4.85184957514486\\
1.04400684931507	-4.87425034332268\\
1.06164383561644	-4.89645113822026\\
1.07928082191781	-4.91845295129444\\
1.09691780821918	-4.94025677400208\\
1.10526847157437	-4.95051154906045\\
1.11455479452055	-4.96186359780003\\
1.13219178082192	-4.98327441414513\\
1.14982876712329	-5.00449021449424\\
1.16746575342466	-5.02551199030421\\
1.17426419331682	-5.03356354917344\\
1.18510273972603	-5.04634073303188\\
1.2027397260274	-5.06697743413411\\
1.22037671232877	-5.08742308506775\\
1.23801369863014	-5.10767867728964\\
1.25565068493151	-5.12774520225664\\
1.27328767123288	-5.14762365142559\\
1.28284203440226	-5.15831407838751\\
1.29092465753425	-5.16731501625335\\
1.30856164383562	-5.18682028819677\\
1.32619863013699	-5.20614045871269\\
1.33960196941994	-5.22069976881568\\
1.34383561643836	-5.22527651925796\\
1.36147260273973	-5.24422946128944\\
1.3791095890411	-5.26300027626397\\
1.39674657534247	-5.28158995563841\\
1.40901924035567	-5.294419201286\\
1.41438356164384	-5.2999994908696\\
1.43202054794521	-5.31822987341439\\
1.44965753424658	-5.33628209472964\\
1.46729452054795	-5.35415714627219\\
1.48493150684932	-5.3718560194989\\
1.50256849315068	-5.38937970586661\\
1.52020547945205	-5.40672919683217\\
1.53784246575342	-5.42390548385243\\
1.55547945205479	-5.44090955838425\\
1.57311643835616	-5.45774241188446\\
1.57646553009519	-5.46091954483484\\
1.59075342465753	-5.47440503580993\\
1.6083904109589	-5.4908984216175\\
1.62602739726027	-5.50722356076403\\
1.64366438356164	-5.52338144470635\\
1.66130136986301	-5.53937306490132\\
1.67893835616438	-5.5551994128058\\
1.69444274006365	-5.56897637593432\\
1.69657534246575	-5.57086147987663\\
1.71421232876712	-5.58636025757066\\
1.73184931506849	-5.60169673734473\\
1.74948630136986	-5.61687191065571\\
1.76712328767123	-5.63188676896044\\
1.7847602739726	-5.64674230371577\\
1.79700955238106	-5.65696657369388\\
1.80239726027397	-5.66143950637854\\
1.82003424657534	-5.67597936840562\\
1.83767123287671	-5.69036288125385\\
1.85530821917808	-5.70459103638007\\
1.87294520547945	-5.71866482524115\\
1.89058219178082	-5.73258523929392\\
1.90821917808219	-5.74635326999524\\
1.92585616438356	-5.75996990880196\\
1.94349315068493	-5.77343614717092\\
1.9611301369863	-5.78675297655899\\
1.97876712328767	-5.799921388423\\
1.99640410958904	-5.81294237421981\\
2.01404109589041	-5.82581692540627\\
2.03167808219178	-5.83854603343922\\
2.04931506849315	-5.85113068977553\\
2.06510176064452	-5.86227338778356\\
2.06695205479452	-5.86357188587202\\
2.08458904109589	-5.87587061318557\\
2.10222602739726	-5.88802786317301\\
2.11986301369863	-5.90004462729119\\
2.1375	-5.91192189699697\\
2.15513698630137	-5.9236606637472\\
2.17277397260274	-5.93526191899872\\
2.19041095890411	-5.94672665420839\\
2.20047243051213	-5.95320623187961\\
2.20804794520548	-5.95805586083306\\
2.22568493150685	-5.96925053032957\\
2.24332191780822	-5.98031165415477\\
2.26095890410959	-5.99124022376552\\
2.27859589041096	-6.00203723061866\\
2.29623287671233	-6.01270366617104\\
2.3138698630137	-6.02324052187952\\
2.33150684931507	-6.03364878920095\\
2.34914383561644	-6.04392945959216\\
2.36678082191781	-6.05408352451002\\
2.38441780821918	-6.06411197541138\\
2.39177270923604	-6.06825711958017\\
2.40205479452055	-6.07401580375307\\
2.41969178082192	-6.08379600099196\\
2.43732876712329	-6.09345355858489\\
2.45496575342466	-6.10298946798871\\
2.47260273972603	-6.11240472066028\\
2.4902397260274	-6.12170030805644\\
2.50787671232877	-6.13087722163404\\
2.52551369863014	-6.13993645284993\\
2.54315068493151	-6.14887899316096\\
2.56078767123288	-6.15770583402398\\
2.57842465753425	-6.16641796689584\\
2.59606164383562	-6.1750163832334\\
2.59739594105384	-6.17566228326089\\
2.61369863013699	-6.18350207449349\\
2.63133561643836	-6.19187603213298\\
2.64897260273973	-6.2001392476087\\
2.6666095890411	-6.20829271237752\\
2.68424657534247	-6.21633741789627\\
2.70188356164384	-6.22427435562182\\
2.71952054794521	-6.232104517011\\
2.73715753424658	-6.23982889352067\\
2.75479452054795	-6.24744847660768\\
2.77243150684932	-6.25496425772888\\
2.79006849315068	-6.26237722834112\\
2.80770547945206	-6.26968837990125\\
2.82534246575342	-6.27689870386612\\
2.84297945205479	-6.28400919169257\\
2.86061643835616	-6.29102083483746\\
2.87825342465753	-6.29793462475764\\
2.8958904109589	-6.30475155290995\\
2.91352739726027	-6.31147261075126\\
2.93116438356164	-6.3180987897384\\
2.93133883546455	-6.3181638603955\\
2.94880136986301	-6.32463108132822\\
2.96643835616438	-6.33107047697759\\
2.98407534246575	-6.33741796814334\\
3.00171232876712	-6.34367454628232\\
3.01934931506849	-6.34984120285139\\
3.03698630136986	-6.3559189293074\\
3.05462328767123	-6.36190871710719\\
3.0722602739726	-6.36781155770762\\
3.08989726027397	-6.37362844256554\\
3.10753424657534	-6.37936036313778\\
3.12517123287671	-6.38500831088122\\
3.14280821917808	-6.39057327725268\\
3.16044520547945	-6.39605625370904\\
3.17808219178082	-6.40145823170712\\
3.19571917808219	-6.40678020270379\\
3.21335616438356	-6.4120231581559\\
3.23099315068493	-6.41718808952029\\
3.2486301369863	-6.42227598825381\\
3.26626712328767	-6.42728784581331\\
3.28390410958904	-6.43222465365565\\
3.29577661200145	-6.43550614521811\\
3.30154109589041	-6.43708740323767\\
3.31917808219178	-6.44187708601622\\
3.33681506849315	-6.44659469344815\\
3.35445205479452	-6.45124121699032\\
3.37208904109589	-6.45581764809957\\
3.38972602739726	-6.46032497823275\\
3.40736301369863	-6.46476419884671\\
3.425	-6.46913630139831\\
3.44263698630137	-6.47344227734438\\
3.46027397260274	-6.47768311814179\\
3.47791095890411	-6.48185981524738\\
3.49554794520548	-6.48597336011799\\
3.51318493150685	-6.49002474421049\\
3.53082191780822	-6.49401495898172\\
3.54845890410959	-6.49794499588853\\
3.56609589041096	-6.50181584638776\\
3.58373287671233	-6.50562850193628\\
3.6013698630137	-6.50938395399093\\
3.61900684931507	-6.51308319400856\\
3.63664383561644	-6.51672721344601\\
3.65428082191781	-6.52031700376014\\
3.67191780821918	-6.52385355640781\\
3.68955479452055	-6.52733786284585\\
3.70719178082192	-6.53077091453112\\
3.72482876712329	-6.53415370292046\\
3.74246575342466	-6.53748721947074\\
3.75861177589174	-6.54049657796109\\
3.76010273972603	-6.54077245563879\\
3.7777397260274	-6.54401040288147\\
3.79537671232877	-6.54720205265563\\
3.81301369863014	-6.55034839641812\\
3.83065068493151	-6.55345042562578\\
3.84828767123288	-6.55650913173547\\
3.86592465753425	-6.55952550620403\\
3.88356164383562	-6.56250054048832\\
3.90119863013699	-6.56543522604519\\
3.91883561643836	-6.56833055433148\\
3.93647260273973	-6.57118751680405\\
3.9541095890411	-6.57400710491974\\
3.97174657534247	-6.57679031013541\\
3.98938356164384	-6.57953812390791\\
4.00702054794521	-6.58225153769408\\
4.02465753424658	-6.58493154295077\\
4.04229452054795	-6.58757913113484\\
4.05993150684932	-6.59019529370313\\
4.07756849315069	-6.5927810221125\\
4.09520547945205	-6.59533730781979\\
4.11284246575342	-6.59786514228185\\
4.13047945205479	-6.60036551695554\\
4.14811643835616	-6.6028394232977\\
4.16575342465753	-6.60528785276518\\
4.1833904109589	-6.60771179681484\\
4.20102739726027	-6.61011224690352\\
4.21866438356164	-6.61249019448807\\
4.22119289042782	-6.61282932007375\\
4.23630136986301	-6.61484663102535\\
4.25393835616438	-6.61718254797219\\
4.27157534246575	-6.61949893678546\\
4.28921232876712	-6.621796788922\\
4.30684931506849	-6.62407709583867\\
4.32448630136986	-6.6263408489923\\
4.34212328767123	-6.62858903983975\\
4.3597602739726	-6.63082265983788\\
4.37739726027397	-6.63304270044352\\
4.39503424657534	-6.63525015311354\\
4.41267123287671	-6.63744600930478\\
4.43030821917808	-6.63963126047408\\
4.44794520547945	-6.64180689807831\\
4.46558219178082	-6.64397391357431\\
4.48321917808219	-6.64613329841892\\
4.50085616438356	-6.64828604406901\\
4.51849315068493	-6.65043314198141\\
4.5361301369863	-6.65257558361299\\
4.55376712328767	-6.65471436042058\\
4.57140410958904	-6.65685046386104\\
4.57169193385716	-6.65688530705099\\
4.58904109589041	-6.65898488539122\\
4.60667808219178	-6.66111861646797\\
4.62431506849315	-6.66325264854813\\
4.64195205479452	-6.66538797308856\\
4.65958904109589	-6.66752558154611\\
4.67722602739726	-6.66966646537763\\
4.69486301369863	-6.67181161603997\\
4.7125	-6.67396202498997\\
4.73013698630137	-6.67611868368449\\
4.74777397260274	-6.67828258358037\\
4.76541095890411	-6.68045471613448\\
4.78304794520548	-6.68263607280364\\
4.80068493150685	-6.68482764504473\\
4.81832191780822	-6.68703042431458\\
4.83595890410959	-6.68924540207004\\
4.85359589041096	-6.69147356976797\\
4.87123287671233	-6.69371591886522\\
4.88442035440476	-6.69540242126539\\
4.8888698630137	-6.69597344081863\\
4.90650684931507	-6.69824712708505\\
4.92414383561644	-6.70053796912135\\
4.94178082191781	-6.70284695838435\\
4.95941780821918	-6.70517508633092\\
4.97705479452055	-6.7075233444179\\
4.99469178082192	-6.70989272410215\\
5.01232876712329	-6.71228421684051\\
5.02996575342466	-6.71469881408983\\
5.04760273972603	-6.71713750730697\\
5.0652397260274	-6.71960128794877\\
5.08287671232877	-6.72209114747208\\
5.10051369863014	-6.72460807733376\\
5.11815068493151	-6.72715306899065\\
5.13578767123288	-6.7297271138996\\
5.14334707003083	-6.73083951740014\\
5.15342465753425	-6.73233120351746\\
5.17106164383562	-6.73496632930108\\
5.18869863013699	-6.73763348270732\\
5.20633561643836	-6.74033365519301\\
5.22397260273973	-6.74306783821502\\
5.2416095890411	-6.74583702323019\\
5.25924657534247	-6.74864220169537\\
5.26534782560261	-6.74962117088711\\
5.27688356164384	-6.75148436506741\\
5.29452054794521	-6.75436450480316\\
5.31215753424658	-6.75728361235947\\
5.32979452054794	-6.7602426791932\\
5.34743150684931	-6.76324269676118\\
5.36506849315068	-6.76628465652027\\
5.38270547945206	-6.76936954992732\\
5.40034246575342	-6.77249836843918\\
5.41797945205479	-6.7756721035127\\
5.42731754097933	-6.77737099605476\\
5.43561643835616	-6.77889174660472\\
5.45325342465753	-6.78215828917211\\
5.4708904109589	-6.78547272267171\\
5.48852739726027	-6.78883603856036\\
5.50616438356164	-6.79224922829492\\
5.52380136986301	-6.79571328333224\\
5.54143835616438	-6.79922919512916\\
5.55907534246575	-6.80279795514255\\
5.57671232876712	-6.80642055482924\\
5.57965585775113	-6.80703046180824\\
5.59434931506849	-6.81009798564609\\
5.61198630136986	-6.81383123904994\\
5.62962328767123	-6.81762130649766\\
5.6472602739726	-6.82146917944608\\
5.66489726027397	-6.82537584935205\\
5.68253424657534	-6.82934230767243\\
5.70017123287671	-6.83336954586407\\
5.71780821917808	-6.83745855538381\\
5.73544520547945	-6.84161032768851\\
5.75308219178082	-6.84582585423501\\
5.77071917808219	-6.85010612648017\\
5.78835616438356	-6.85445213588083\\
5.80599315068493	-6.85886487389385\\
5.8110436127388	-6.86014091831871\\
5.8236301369863	-6.86334533197607\\
5.84126712328767	-6.86789450158435\\
5.85890410958904	-6.87251337417553\\
5.87654109589041	-6.87720294120647\\
5.89417808219178	-6.88196419413401\\
5.91181506849315	-6.886798124415\\
5.92945205479452	-6.89170572350629\\
5.94708904109589	-6.89668798286474\\
5.95359933305229	-6.89854613430779\\
5.96472602739726	-6.90174589394719\\
5.98236301369863	-6.9068804482105\\
6	-6.91209263711151\\
};
\addlegendentry{\small{Proposed scheme}}

\addplot [color=red, dashed, line width=1.0pt]
  table[row sep=crcr]{%
1.82354091089179	1.19740541200162\\
1.9	0.851972808937781\\
1.9012163615359	0.846559705110513\\
1.91409726027397	0.789393741827734\\
1.92819452054795	0.727158058586834\\
1.94229178082192	0.66526479077023\\
1.95638904109589	0.603712969933062\\
1.97048630136986	0.542501627630477\\
1.98318991889624	0.487632435733564\\
1.98458356164384	0.481629795417614\\
1.99868082191781	0.421096504849626\\
2.01277808219178	0.360900787481645\\
2.02687534246575	0.301041674868827\\
2.04097260273973	0.241518198566308\\
2.0550698630137	0.182329390129237\\
2.06916712328767	0.123474281112756\\
2.08326438356164	0.0649519030720036\\
2.08733530440183	0.0481139454828465\\
2.09736164383562	0.00676128756213146\\
2.11145890410959	-0.0510985338617189\\
2.12555616438356	-0.108628529644404\\
2.12885592096199	-0.122047087562439\\
2.13965342465753	-0.16582966823078\\
2.15375068493151	-0.222702918065703\\
2.16000725848124	-0.247839310714959\\
2.16784794520548	-0.279249247594029\\
2.18194520547945	-0.335469625260611\\
2.19604246575342	-0.39136501951031\\
2.2101397260274	-0.44693639878798\\
2.22423698630137	-0.502184731538474\\
2.23833424657534	-0.55711098620665\\
2.25243150684932	-0.611716131237367\\
2.26652876712329	-0.666001135075479\\
2.28062602739726	-0.719966966165838\\
2.29472328767123	-0.773614592953308\\
2.30882054794521	-0.826944983882738\\
2.32291780821918	-0.879959107398987\\
2.33190636651553	-0.913596786470434\\
2.33701506849315	-0.932657931946913\\
2.35111232876712	-0.985042425971365\\
2.3652095890411	-1.03711355791721\\
2.37930684931507	-1.08887229622929\\
2.39340410958904	-1.14031960935247\\
2.40750136986301	-1.19145646573161\\
2.42159863013699	-1.24228383381156\\
2.43489585030199	-1.28994390447652\\
2.43569589041096	-1.29280268203717\\
2.44979315068493	-1.34301397885331\\
2.4638904109589	-1.39291869270482\\
2.47798767123288	-1.44251779203658\\
2.49208493150685	-1.49181224529341\\
2.50618219178082	-1.5408030209202\\
2.52027945205479	-1.58949108736179\\
2.53437671232877	-1.63787741306304\\
2.54847397260274	-1.6859629664688\\
2.56257123287671	-1.73374871602394\\
2.57652399299829	-1.78075039101609\\
2.57666849315068	-1.7812356301733\\
2.59076575342466	-1.82842467736174\\
2.60486301369863	-1.87531682603412\\
2.6189602739726	-1.9219130446353\\
2.63305753424658	-1.96821430161013\\
2.64715479452055	-2.01422156540347\\
2.66125205479452	-2.05993580446016\\
2.67534931506849	-2.10535798722508\\
2.67562068027904	-2.10622948284475\\
2.68944657534247	-2.15048908214307\\
2.70354383561644	-2.19533005765899\\
2.71764109589041	-2.2398818822177\\
2.73173835616438	-2.28414552426405\\
2.74583561643836	-2.3281219522429\\
2.75993287671233	-2.37181213459911\\
2.7740301369863	-2.41521703977752\\
2.78812739726027	-2.458337636223\\
2.80222465753425	-2.50117489238041\\
2.80509939456037	-2.50987564420752\\
2.81632191780822	-2.54372977669459\\
2.83041917808219	-2.58600325761041\\
2.84451643835616	-2.62799630357272\\
2.85861369863014	-2.66970988302638\\
2.87271095890411	-2.71114496441624\\
2.88680821917808	-2.75230251618715\\
2.90090547945205	-2.79318350678399\\
2.91500273972603	-2.83378890465159\\
2.9291	-2.87411967823482\\
2.94319726027397	-2.91417679597854\\
2.95729452054795	-2.95396122632759\\
2.97139178082192	-2.99347393772684\\
2.98548904109589	-3.03271589862114\\
2.99958630136986	-3.07168807745534\\
3.01368356164384	-3.11039144267431\\
3.02131714978317	-3.13123725158808\\
3.02778082191781	-3.1488269627229\\
3.04187808219178	-3.18699560604596\\
3.05597534246575	-3.22489834108835\\
3.07007260273973	-3.26253613629493\\
3.0841698630137	-3.29990996011055\\
3.09826712328767	-3.33702078098008\\
3.11236438356164	-3.37386956734835\\
3.12646164383562	-3.41045728766024\\
3.14055890410959	-3.4467849103606\\
3.15266757966406	-3.47778117988423\\
3.15465616438356	-3.48285340389427\\
3.16875342465753	-3.51866373670613\\
3.18285068493151	-3.55421687724102\\
3.19694794520548	-3.58951379394381\\
3.21104520547945	-3.62455545525933\\
3.22514246575342	-3.65934282963246\\
3.2392397260274	-3.69387688550806\\
3.25333698630137	-3.72815859133096\\
3.26743424657534	-3.76218891554604\\
3.28153150684932	-3.79596882659814\\
3.29562876712329	-3.82949929293213\\
3.30972602739726	-3.86278128299286\\
3.32382328767123	-3.89581576522518\\
3.33792054794521	-3.92860370807395\\
3.35201780821918	-3.96114607998403\\
3.35685684376248	-3.97226015701775\\
3.36611506849315	-3.99344384940028\\
3.38021232876712	-4.02549798476754\\
3.3943095890411	-4.05730945453067\\
3.40840684931507	-4.08887922713454\\
3.42250410958904	-4.120208271024\\
3.43660136986301	-4.1512975546439\\
3.45069863013699	-4.1821480464391\\
3.46479589041096	-4.21276071485445\\
3.47889315068493	-4.24313652833481\\
3.47983949243154	-4.24516718113695\\
3.4929904109589	-4.27327645532505\\
3.50708767123288	-4.30318146427\\
3.52118493150685	-4.33285252361453\\
3.53528219178082	-4.3622906018035\\
3.54937945205479	-4.39149666728176\\
3.56347671232877	-4.42047168849417\\
3.57757397260274	-4.44921663388557\\
3.59167123287671	-4.47773247190084\\
3.60576849315069	-4.50602017098483\\
3.61986575342466	-4.53408069958238\\
3.63396301369863	-4.56191502613837\\
3.6480602739726	-4.58952411909763\\
3.65525893764437	-4.6035359422315\\
3.66215753424658	-4.61690894690504\\
3.67625479452055	-4.64407047800544\\
3.69035205479452	-4.67100968084369\\
3.70444931506849	-4.69772752386465\\
3.71854657534247	-4.72422497551317\\
3.73264383561644	-4.75050300423411\\
3.74674109589041	-4.77656257847233\\
3.76083835616438	-4.80240466667268\\
3.77493561643836	-4.82803023728001\\
3.78903287671233	-4.85344025873919\\
3.8031301369863	-4.87863569949506\\
3.81722739726027	-4.90361752799249\\
3.83132465753425	-4.92838671267634\\
3.8368240849329	-4.93799187503242\\
3.84542191780822	-4.95294422199145\\
3.85951917808219	-4.97729102438268\\
3.87361643835616	-5.0014280882949\\
3.88771369863014	-5.02535638217295\\
3.90181095890411	-5.04907687446169\\
3.91590821917808	-5.07259053360598\\
3.93000547945205	-5.09589832805067\\
3.94410273972603	-5.11900122624062\\
3.9582	-5.14190019662068\\
3.97229726027397	-5.16459620763572\\
3.98639452054795	-5.18709022773058\\
4.00049178082192	-5.20938322535013\\
4.01458904109589	-5.23147616893922\\
4.02238399782767	-5.24360673091371\\
4.02868630136986	-5.2533700269427\\
4.04278356164384	-5.27506576780543\\
4.05688082191781	-5.29656435997227\\
4.07097808219178	-5.31786677188808\\
4.08507534246575	-5.3389739719977\\
4.09917260273973	-5.359886928746\\
4.1132698630137	-5.38060661057784\\
4.12736712328767	-5.40113398593805\\
4.14146438356164	-5.42147002327151\\
4.15556164383562	-5.44161569102307\\
4.16965890410959	-5.46157195763758\\
4.18375616438356	-5.48133979155992\\
4.19785342465753	-5.50092016123491\\
4.21195068493151	-5.52031403510743\\
4.22604794520548	-5.53952238162233\\
4.24014520547945	-5.55854616922446\\
4.25424246575342	-5.57738636635869\\
4.2683397260274	-5.59604394146986\\
4.28243698630137	-5.61451986300283\\
4.29349757932355	-5.6288893730329\\
4.29653424657534	-5.63281509940247\\
4.31063150684932	-5.65093061911362\\
4.32472876712329	-5.66886739058114\\
4.33882602739726	-5.68662638224989\\
4.35292328767123	-5.70420856256473\\
4.36702054794521	-5.7216148999705\\
4.38111780821918	-5.73884636291207\\
4.39521506849315	-5.75590391983429\\
4.40931232876712	-5.77278853918202\\
4.4234095890411	-5.78950118940011\\
4.42361181780061	-5.78973969015524\\
4.43750684931507	-5.80604283893342\\
4.45160410958904	-5.82241445622681\\
4.46570136986301	-5.83861700972512\\
4.47979863013699	-5.85465146787323\\
4.49389589041096	-5.87051879911598\\
4.50799315068493	-5.88621997189822\\
4.5220904109589	-5.90175595466483\\
4.53618767123288	-5.91712771586064\\
4.55028493150685	-5.93233622393053\\
4.56438219178082	-5.94738244731933\\
4.5784794520548	-5.96226735447192\\
4.59257671232877	-5.97699191383315\\
4.60667397260274	-5.99155709384786\\
4.62077123287671	-6.00596386296092\\
4.63407100519848	-6.01941128383268\\
4.63486849315069	-6.02021318961718\\
4.64896575342466	-6.03430604226151\\
4.66306301369863	-6.04824338933875\\
4.6771602739726	-6.06202619929376\\
4.69125753424658	-6.0756554405714\\
4.70535479452055	-6.08913208161653\\
4.71945205479452	-6.10245709087399\\
4.73354931506849	-6.11563143678865\\
4.74764657534247	-6.12865608780536\\
4.76174383561644	-6.14153201236898\\
4.77584109589041	-6.15426017892436\\
4.78993835616438	-6.16684155591636\\
4.80403561643836	-6.17927711178984\\
4.81813287671233	-6.19156781498964\\
4.8322301369863	-6.20371463396064\\
4.84632739726027	-6.21571853714768\\
4.86042465753425	-6.22758049299562\\
4.87136774785384	-6.23669117085117\\
4.87452191780822	-6.23930146994932\\
4.88861917808219	-6.25088243645363\\
4.90271643835616	-6.2623243609534\\
4.91681369863014	-6.2736282118935\\
4.93091095890411	-6.28479495771878\\
4.94500821917808	-6.29582556687409\\
4.95910547945206	-6.3067210078043\\
4.97320273972603	-6.31748224895426\\
4.9873	-6.32811025876882\\
4.99281381728739	-6.33223112292071\\
5.00139726027397	-6.33860600569283\\
5.01549452054795	-6.34897045817117\\
5.02959178082192	-6.35920458464867\\
5.04368904109589	-6.3693093535702\\
5.05778630136986	-6.37928573338062\\
5.07188356164384	-6.38913469252478\\
5.08598082191781	-6.39885719944753\\
5.10007808219178	-6.40845422259373\\
5.11417534246575	-6.41792673040824\\
5.12827260273973	-6.42727569133592\\
5.1423698630137	-6.43650207382162\\
5.15646712328767	-6.44560684631019\\
5.17056438356164	-6.4545909772465\\
5.18466164383562	-6.46345543507539\\
5.19434471374659	-6.46947538991085\\
5.19875890410959	-6.47220118824172\\
5.21285616438356	-6.48082920519036\\
5.22695342465753	-6.48934045436615\\
5.24105068493151	-6.49773590421396\\
5.25514794520548	-6.50601652317863\\
5.26924520547945	-6.51418327970502\\
5.28334246575342	-6.522237142238\\
5.2974397260274	-6.53017907922241\\
5.31153698630137	-6.53801005910311\\
5.32563424657534	-6.54573105032496\\
5.33973150684932	-6.55334302133282\\
5.35382876712329	-6.56084694057153\\
5.36792602739726	-6.56824377648596\\
5.37865808394189	-6.57380369576236\\
5.38202328767123	-6.57553449752097\\
5.39612054794521	-6.5827200721214\\
5.41021780821918	-6.58980146873211\\
5.42431506849315	-6.59677965579796\\
5.43841232876712	-6.60365560176382\\
5.4525095890411	-6.61043027507452\\
5.46660684931507	-6.61710464417493\\
5.48070410958904	-6.6236796775099\\
5.49480136986301	-6.6301563435243\\
5.50889863013699	-6.63653561066297\\
5.52299589041096	-6.64281844737078\\
5.53709315068493	-6.64900582209257\\
5.5511904109589	-6.6550987032732\\
5.56528767123288	-6.66109805935754\\
5.57938493150685	-6.66700485879043\\
5.59348219178082	-6.67282007001673\\
5.60757945205479	-6.6785446614813\\
5.62167671232877	-6.684179601629\\
5.63577397260274	-6.68972585890467\\
5.63740106851146	-6.69036033943632\\
5.64987123287671	-6.69518440175319\\
5.66396849315069	-6.70055619861939\\
5.67806575342466	-6.70584221794814\\
5.69216301369863	-6.7110434281843\\
5.7062602739726	-6.71616079777271\\
5.72035753424658	-6.72119529515825\\
5.73445479452055	-6.72614788878575\\
5.74855205479452	-6.73101954710008\\
5.76264931506849	-6.7358112385461\\
5.77674657534247	-6.74052393156866\\
5.79084383561644	-6.74515859461261\\
5.80494109589041	-6.74971619612282\\
5.81670232922919	-6.75346030229808\\
5.81903835616438	-6.75419770454413\\
5.83313561643836	-6.75860408832141\\
5.84723287671233	-6.7629363158995\\
5.8613301369863	-6.76719535572328\\
5.87542739726027	-6.77138217623758\\
5.88952465753425	-6.77549774588728\\
5.90362191780822	-6.77954303311721\\
5.91771917808219	-6.78351900637226\\
5.93181643835617	-6.78742663409725\\
5.94591369863014	-6.79126688473706\\
5.96001095890411	-6.79504072673653\\
5.97410821917808	-6.79874912854053\\
5.98820547945206	-6.80239305859391\\
6.00230273972603	-6.80597348534153\\
6.0164	-6.80949137722824\\
};
\addlegendentry{\small{Clipped RZF}}

\addplot [color=red, draw=none, mark size=3.0pt, mark=o, mark options={solid, red}, forget plot]
  table[row sep=crcr]{%
1.82354091089179	1.25757122339784\\
1.98318991889624	0.518290321838211\\
2.12885592096199	-0.124745774610862\\
2.33190636651553	-1.02110396046365\\
2.57652399299829	-1.73540894579278\\
2.80509939456037	-2.56239113006664\\
3.02131714978317	-3.10663535472137\\
3.15266757966406	-3.46299463556856\\
3.35685684376248	-3.84419095865347\\
3.47983949243154	-4.22931290376892\\
3.65525893764437	-4.54384061412294\\
3.8368240849329	-4.93072239640694\\
4.02238399782767	-5.25357546808593\\
4.29349757932355	-5.57078205256302\\
4.63407100519848	-6.04635822798689\\
4.87136774785384	-6.27009441737223\\
5.37865808394189	-6.65238222619958\\
5.81670232922919	-6.73495600255744\\
};

\addplot [color=mycolor1, draw=none, mark size=3.0pt, mark=o, mark options={solid, mycolor1},forget plot]
  table[row sep=crcr]{%
0.856686385900919	-4.53532181968934\\
0.955945551252391	-4.73170715416063\\
1.17426419331682	-5.02937618895169\\
1.33960196941994	-5.30087076150997\\
1.69444274006365	-5.64490270652331\\
2.39177270923604	-6.05404448133452\\
3.29577661200145	-6.36628986988536\\
4.22119289042782	-6.65554931288377\\
4.57169193385716	-6.65247544741582\\
4.88442035440476	-6.71648869962701\\
5.26534782560261	-6.72186451848526\\
5.95359933305229	-6.84645236272709\\
};

\end{axis}
\end{tikzpicture}%